\documentclass[12pt]{spieman}  
\usepackage{amsmath,amsfonts,amssymb}
\usepackage{graphicx}
\usepackage{adjustbox}
\usepackage{setspace}
\usepackage{tocloft}
\usepackage{lineno}

\title{Daytime turbulence strength profile measurement at Kodaikanal Observatory}

\author[a,b]{Saraswathi Kalyani Subramanian *}
\author[a]{Sridharan Rengaswamy}
\author[a]{Prasanna Gajanan Deshmukh}
\author[a]{Binukumar G. Nair}
\author[a]{Mahesh Babu S}

\affil[a]{Indian Institute of Astrophysics, 100 Feet Road, Santhoshapuram, 2nd Block Koramangala, Bengaluru, India}
\affil[b]{Department of Applied Optics and Photonics, University of Calcutta,  87/1, College Street, Kolkata, India}

\cftpagenumbersoff{figure}
\cftpagenumbersoff{table} 
\begin{document} 
\maketitle
\begin{abstract}
The Indian Institute of Astrophysics (IIA) is developing a Multi-Conjugate Adaptive Optics (MCAO) system for the Kodaikanal Tower Telescope (KTT). In this context, we have measured the daytime turbulence strength profile at the Kodaikanal Observatory. The first method based on wavefront sensor (WFS) images, called S-DIMM+ (Solar-Differential Image Motion Monitor+), was used to estimate the higher altitude turbulence up to a height of 5 - 6~km. The second method used balloon-borne temperature sensors to measure the near-Earth turbulence up to 350~m. We also carried out simulations to validate the performance of our system. We report the first-ever daytime turbulence strength profile measurements at the observatory. We have identified the presence of a strong turbulence layer about 3~km above the observatory. The measured near-Earth turbulence matches the trend that is expected from the model for daytime component of turbulence and gives an integrated $r_0$ of about 4~cm at 500~nm. This is consistent with earlier seeing measurements. This shows that a low-cost setup with a small telescope and a simple array of temperature sensors can be used for estimating the turbulence strength profile at the site.  

\end{abstract}


{\noindent \footnotesize\textbf{*}Saraswathi Kalyani Subramanian,  \linkable{saraswathi.kalyani@iiap.res.in} }

\begin{spacing}{2}   

\section{Introduction}
\label{sect:intro}  

Babcock \cite{1953PASP...65..229B} put forth the idea of compensating for the effects of atmospheric turbulence on the light coming from celestial bodies using Adaptive Optics (AO) in 1953. Since then, the field of AO has grown tremendously with the advent of various types of AO systems based on the scientific goal. An essential aspect in ascertaining a new telescope site\cite{1986ESOC...24..229S, 1988SoPh..115..183Z, 2005PASP..117.1296S} or in developing an AO system is the distribution of the turbulence strength profile at that site\cite{2016SPIE.9909E..6XM}. Additionally, seeing measurements are routinely carried out at observatories to improve efficiency in scheduling observations. 
Therefore, there are many reasons why quantifying the turbulence strength at telescope sites is essential. There has been much research in this area with many methods used to quantify the site characteristics \cite{1985ARA&A..23...19C}. 

Some of the earliest turbulence profile measurements were done by using temperature sensors to measure the temperature structure function ($D_T(r)$) and then estimating the turbulence strength profile ($C_N^2(h)$) using the relationship between it and temperature structure function parameter ($C_T^2$). Typically, the sensors are hoisted using a mechanism like a balloon. \cite{1972JOSA...62.1068B, 1976JOSA...66.1380B}. Hereafter, we refer to this method as balloon-measurements in this paper.

Other methods were developed to estimate site characteristics from telescope images. The first detailed study with a differential image motion monitor (DIMM) at ESO was published by Sarazin and Roddier \cite{1990A&A...227..294S}. DIMM uses the differential image motions of the same object as seen by two apertures mounted on a common platform to determine the Fried’s parameter ($r_0$)\cite{1987PASP...99.1360M, 2002PASP..114.1156T}. The use of the “differential” motion minimizes the effect of telescope jitter and tracking errors on $r_0$ estimation. The concept was later extended to daytime measurements by observing the sun and is known as S-DIMM (Solar DIMM) \cite{2001ExA....12....1B, 2011MNRAS.416.2154K}, where the image motion of the solar limb is typically measured. While the (S-)DIMM is a good instrument in determining the statistics of $r_0$ at the site, it does not provide information on the distribution of the strength of turbulence as a function of height. If the $C_N^2(h)$ profile can be found, the $r_0$ value can be calculated from the measurements as its integral\cite{1981PrOpt..19..281R}. In 1993, Seykora\cite{1993SoPh..145..389S} showed that scintillation measurements can be used to estimate the $C_N^2(h)$ profile. It was further extended by Beckers \cite{1993SoPh..145..399B, 1999ASPC..184..309B, 2001SoPh..198..197L} and is known as SHABAR (SHAdow BAnd Ranging). It was extensively used in the DKIST (Daniel K. Inouye Solar Telescope, formerly Advanced Technology Solar Telescope, ATST) site survey\cite{2005PASP..117.1296S}. However, it requires long baselines to estimate high altitude turbulence. As reported in Hickson and Lanzetta\cite{2004PASP..116.1143H}, a baseline of 2.6~m would allow measurement of turbulence profile only up to a height of 300~m. Furthermore, there are challenges in maintaining the pointing of the array and tracking the source.


Later on, there were developments of instruments using an SHWFS (Shack-Hartman Wave Front Sensor) to measure high-altitude seeing using the differential image motions of the sub-aperture images \cite{Waldmann:07}. In 2010, Scharmer and van Werkhoven\cite{2010A&A...513A..25S} proposed the S-DIMM+ method (similar to SLODAR \cite{2002MNRAS.337..103W} for nighttime), which uses a WFS to capture images of a region on the sun. Then, the covariances of image motion of the different sub-aperture images are used to determine the $C_N^2(h)$ profile. It was later repeated with modifications at the Big Bear Solar Observatory (BSSO)\cite{2012A&A...542A...2K} and Fuxian Solar Observatory (FSO)\cite{2018MNRAS.478.1459W}. The measurements were all made with telescopes of apertures 1~m or more since large telescope apertures are required to measure the high altitude turbulence. 
This is because the largest separation between the lenslets (as projected on the pupil plane) determines the maximum height for which inversion is possible.
Further improvements include the work by Ren et al. \cite{2015PASP..127..870R} called Multi Aperture Seeing Profiler (MASP), using a combination of two smaller telescopes to achieve the performance of a single larger telescope and by Ran et al. \cite{2024MNRAS.528.3981R} using autocorrelation in the place of cross-correlation.


Owing to the low cost and simplicity of the setup, we have implemented the S-DIMM+ proposed by Scharmer and van Werkhoven to measure the daytime turbulence profile at Kodaikanal Solar Observatory up to a height of about 5 - 6~km. We have also used balloon-measurements to measure the near-Earth turbulence near-simultaneously. The paper is organized as follows. In section \ref{sec:S_DIMM} we briefly recap the theory of operation of S-DIMM+ along with the simulation procedure we followed to test the inversion code developed by us. Section \ref{sec:balloon_theory} discusses the principle of using temperature sensors to measure temperature fluctuations and, consequently, the turbulence strength profile. Section \ref{sec:expt_obs} describes the two observational setups we have used, the data we acquired, and their analysis. We summarise and discuss our results and future scope of the work in section \ref{sec:result}.

\section{Turbulence characterization with S-DIMM+}
\label{sec:S_DIMM}

\subsection{Principle of the method}
\label{sec:SDIMMplusprinciple}

Scharmer and van Werkhoven \cite{2010A&A...513A..25S} extended the work of Fried \cite{1975RaSc...10...71F} and used covariances of differential image motions to measure the $C_N^2$ profile. While the traditional DIMM uses only two images, S-DIMM+ produces multiple sub-images of the same region of the sun using a SHWFS. Then, the longitudinal ($\delta_{x1}$) and transverse ($\delta_{y1}$) image motions measured using one subfield of two sub-apertures, and the longitudinal ($\delta_{x2}$) and transverse ($\delta_{y2}$) image motions measured using a different subfield of the same two sub-apertures are estimated. It is repeated for all combinations of sub-apertures and sub-fields to build the longitudinal and transverse covariance matrices. The Fourier transform-based cross-correlation method is used to estimate the image motions. The time-averaged covariances of the two sets of image motion measurements ($\delta_{x1}$, $\delta_{x2}$, $\delta_{y1}$, $\delta_{y2}$) are expressed as a sum of seeing contributions from different layers of the atmosphere as\cite{2010A&A...513A..25S} :



\begin{equation}
\label{eq:long_cov}
    \langle \delta_{x1} \delta_{x2}\rangle = \sum_{n=1}^N c_n F_x(s,\alpha, h_n),
\end{equation}
and 
\begin{equation}
\label{eq:tran_cov}
    \langle \delta_{y1} \delta_{y2}\rangle = \sum_{n=1}^N c_n F_y(s,\alpha, h_n).
\end{equation}

The functions $F_x$ and $F_y$ are given by:
\begin{equation}
\label{eq:Fx}
    F_x(s, \alpha, h_n) = 0.5*I\left(\frac{\alpha h_n - s}{D_{\text{eff}}},0 \right)
    + 0.5*I\left(\frac{\alpha h_n + s}{D_{\text{eff}}},0 \right) - I\left(\frac{\alpha h_n}{D_{\text{eff}}},0 \right),
\end{equation}
\begin{equation}
\label{eq:Fy}
    F_y(s, \alpha, h_n) = 0.5*I\left(\frac{\alpha h_n - s}{D_{\text{eff}}},\frac{\pi}{2} \right)
    + 0.5*I\left(\frac{\alpha h_n + s}{D_{\text{eff}}},\frac{\pi}{2} \right) - I\left(\frac{\alpha h_n}{D_{\text{eff}}},\frac{\pi}{2} \right),
\end{equation}

where the function $I$ is given by Fried\cite{1975RaSc...10...71F}. $F_x$ and $F_y$ are a function of the variables `$s$’ and $\alpha$, the linear and angular separation between the regions used for image motion measurements in the pupil and image planes, respectively, and $h_n$ - the height above the ground in the Earth’s atmosphere that is being probed. 
$D_{\text{eff}}$, is the effective diameter of the sub-pupil projected at different heights. These two functions ($F_x$ and $F_y$ ) can be estimated using the system parameters. The left-hand side of equations \ref{eq:long_cov} and \ref{eq:tran_cov} are the measured longitudinal and transverse covariance matrices, respectively. They are calculated from the sub-aperture images (simulated or observed). Each element of the matrices corresponds to two sub-fields having a combination of the linear pupil plane (s) and angular image plane ($\alpha$) separations which were used to determine the image motions. By iterating over all possible combinations of `s' and $\alpha$, the two matrices can be constructed. One frame of lenslet array images produces one longitudinal and one transverse covariance matrix. A time series of such frames is used for determining two matrices for each instant, which are then used to find the ensemble average. The equations are then solved for the $c_n$ coefficients using a linear least squares fit \cite{2010A&A...513A..25S}. Then, using equations \ref{eq:cntor0} and \ref{eq:cntoCn}, we can determine the $r_0$ values and the turbulent strengths of the layers ($C_N^2dh$) at different heights, respectively. 

\begin{equation}
    \label{eq:cntor0}
    c_n = 0.358 \lambda^2 r_0^{-5/3}(h_n)D_{\text{eff}}^{-1/3}(h_n),
\end{equation}
or
\begin{equation}
    \label{eq:cntoCn}
    c_n = 5.98D^{-1/3}_{\text{eff}}(h_n)C_N^2dh/cos(z),
\end{equation}

where $\lambda$ is the wavelength of observation and $cos(z)$ is the inverse of the airmass. The reader is referred to original work\cite{2010A&A...513A..25S} for a more detailed derivation of the above equations. 

\subsection{Simulation procedure}
\label{sec:sim_procedure}
We developed our own inversion code in Python, which uses the covariance of image motions from a sequence of instantaneous WFS images and inverts them to obtain the turbulence strength profile (Eq:\ref{eq:long_cov} and Eq:\ref{eq:tran_cov}). The preliminary version of the code was reported in Subramanian and Rengaswamy\cite{2023SPIE12638E..12S}. Here, we repeat the details of the simulation procedure for the reader’s ease. It must also be noted that some parameters of our simulation between our earlier work and this paper have changed. This was a design choice, and it is explained in section \ref{sec:expt_obs}. However, this shows that our code is robust and is able to invert and retrieve the turbulence strength for different configurations. Furthermore, here, we have used our own code to generate phase screens that describe the phase perturbations in the atmosphere using Kolmogorov's model of turbulence\cite{2004SPIE.5171..219S}. 

The primary goal of the simulation was to test if the S-DIMM+ method can be applied to a smaller telescope with our system parameters. Furthermore, it was also used to determine the optimal size of the sub-field used for correlation ($\phi$), the number of sub-apertures to be used for the inversion, the number of temporal averages required and the best height grid for which our system can perform the inversion. We also verified its ability to retrieve the given input turbulence strength profile successfully.

\begin{figure}[H]
\begin{center}
\begin{tabular}{c}
\includegraphics[height=8cm]{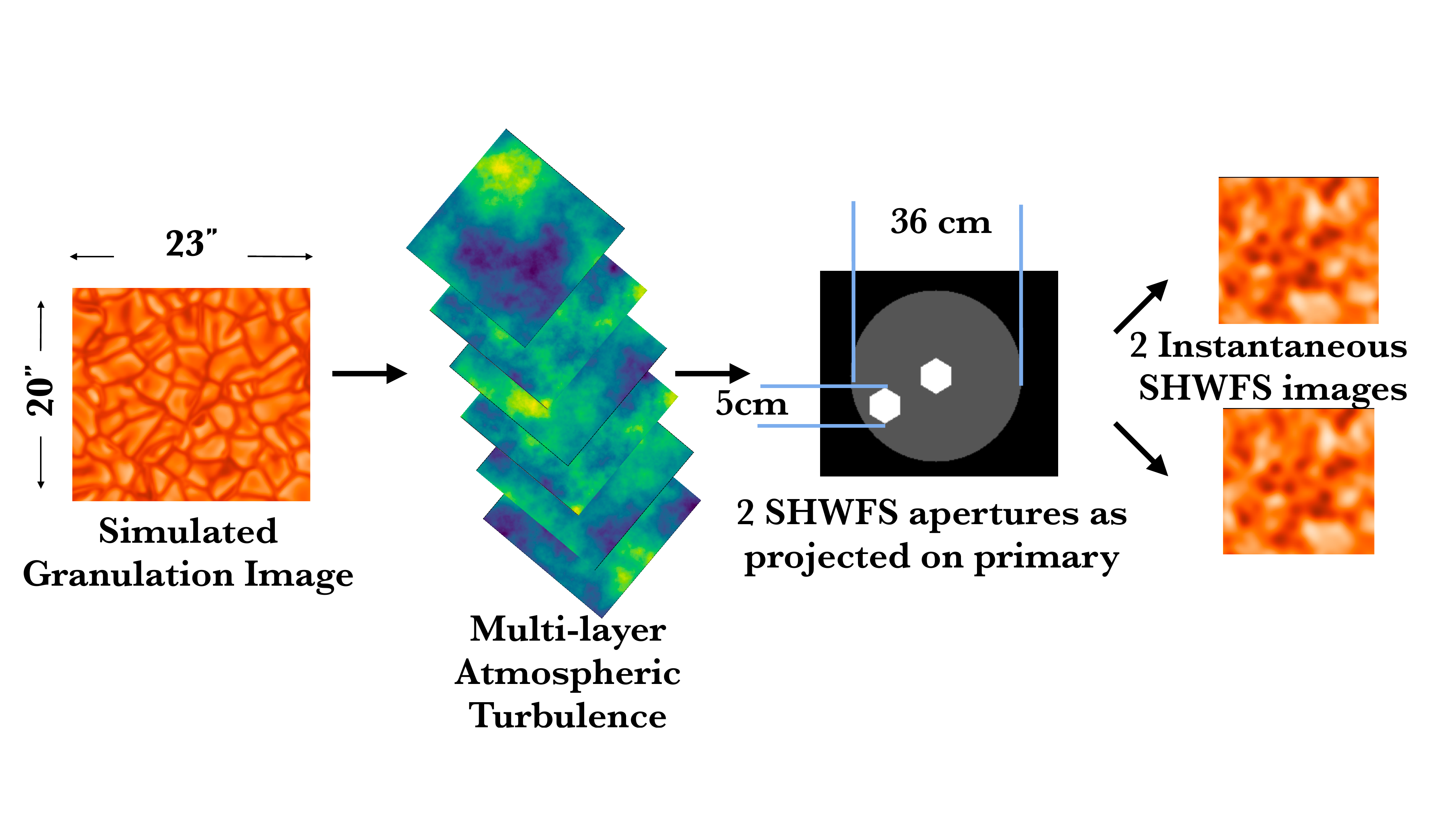}
\end{tabular}
\end{center}
\caption 
{ \label{fig:sim_flow} 
Diagrammatic representation of simulation workflow (images not scaled). From left to right are the simulated solar granulation used as “objects”, followed by multi-layer atmospheric turbulence, a telescope with a lenslet array, and finally, two sets of instantaneous images produced by the SHWFS. (Adapted from Fig 1 (a) in \cite{2023SPIE12638E..12S}.)} 
\end{figure}

As input to the inversion code, we produced a set of instantaneous images that were shifted due to multi-layer atmospheric turbulence, similar to our experimental setup. Figure~\ref{fig:sim_flow} shows the simulation procedure diagrammatically.

The steps involved in simulating WFS images were:
\begin{itemize}
    \item Choosing an ``object''. We used simulated solar granulation images of about 23$''$$\times$20$''$ to match the images we got from our experiment. 
    \item Generating multiple phase screens that follow Kolmogorov’s theory of atmospheric turbulence with the variation in projected pupil size with height accounted for.
    \item Determining the cumulative Optical Transfer Function (OTF) of each lenslet as the product of the OTFs of each individual layer by considering the portion of the phase screen that that lenslet would have sampled. 
    \item Finding the instantaneous image of each lenslet as the inverse Fourier transform of the product obtained by multiplying the cumulative OTF of each lenslet with the Fourier transform of the solar “object”. 
\end{itemize}
 
The above steps produce one frame of instantaneous images. The process is repeated by sampling different regions of the phase screens to produce a series of frames similar to observations which are used as input for the inversion code. The simulations were done at H-$\alpha$ wavelength (656.3~nm). We found that using 200 frames of lenslet array images, which correspondingly produce 200 longitudinal and transverse covariance matrices each, is sufficient for the ensemble average.

A point to be noted is that our simulations assume that the PSF (at a given instant of time) for one sub-aperture is uniform across the field-of-view. This is not true in reality; the PSF is uniform only within the isoplanatic angle. During the daytime, and for seeing-limited imaging, the isoplanatic angle is larger than that for the diffraction-limited case\cite{1975OptCo..14..200W}. It can be as large as 17$''$ \cite{2023aoel.confE..22K}. The field-of-view of each sub-aperture image in our system is 23$''\times$20$''$. Therefore, our assumption of a uniform PSF is not a gross violation of reality.

We highlight through our simulations that improper sampling of the phase screen leads to a systematic underestimation of $r_0$ or a failure of the inversion code. Therefore, it was always ensured that the error in the sub-pupil area (due to discrete sampling) was always less than 2\%. 

We also verified the primary assumption (Eq. \ref{eq:long_cov} and Eq \ref{eq:tran_cov}) which states that the image motions due to multiple thin layers can be expressed as a sum of the individual image motions of each of the layers. During this process, we identified that the cumulative image motion (the parameter measured from the experiment) can be underestimated due to the finite size of the window we are using for the image motion estimation, which in turn can cause incorrect inversions. Therefore, choosing the right size of sub-field is essential. A smaller sub-field can lead to inaccurate image motion estimates which causes a failure of the inversion. On the contrary, it allows a larger maximum angular separation between two sub-fields which decreases the minimum height to which the system is sensitive. The size of the sub-field is also inversely proportional to the maximum height up to which the system is sensitive. The trade-off between height sensitivity and accuracy was studied by varying the size of the sub-field chosen for analysis. We found that using a 13.2$''\times$13.2$''$ sub-field allowed at least 180 of every 200 frames to have accurate image motion measurements (the sum of individual image motions matched the cumulative image motion). This was set as a threshold to identify instances of poor seeing where the magnitude of image motion is very high, increasing the likelihood of underestimation. Therefore, for every batch of 200 frames, it was ensured that less than 10 \% of the frames were ``bad'' with a bad frame being defined as one that returned image motion values within 6 pixels of the edge on any side. If more than 10 \% of the frames were bad, then all 200 frames were rejected from the analysis (simulation and real data).

\subsection{Inversion code}
\label{sec:inv_code}
The inversion code takes a series of instantaneous wavefront sensor images, estimates the longitudinal and transverse covariance matrices, and then fits them to the sum of theoretical matrices at different heights to estimate the turbulence strength profile. We found that using only the sub-aperture images along the horizontal diameter of the pupil was sufficient to cover all the unique separations and estimate the given input turbulence. As stated in the preceding section, a 32$\times$32 window (about 13.2$\times$13.2 arc-seconds$^2$) allowed us to estimate the image motions accurately and retrieve the given input turbulence parameters.
We have used 12 unique values of angular separations from 0 to about 9$''$ in steps of about 0.8$''$. For our system parameters, the minimum and maximum heights are 0~km (pupil of the telescope) and 6~km, respectively. We also found that a height grid with layers roughly 0, 1, 2, 3, 4, and 6~km above the ground gave good inversions able to detect and identify the presence of a strong layer of turbulence). It must be noted that the “ground layer” in the S-DIMM+ procedure is the height of the primary mirror, which is at 2345.38~m above sea level. For the remainder of the paper it will be referred to as GL-(S-DIMM+).

\begin{figure}[H]
\begin{center}
\begin{tabular}{c}
\includegraphics[height=8cm]{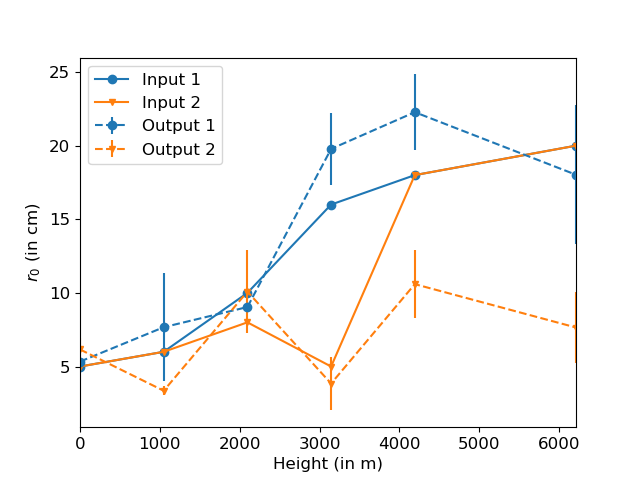}
\end{tabular}
\end{center}
\caption 
{ \label{fig:s_dimm_sim_res} Results of the inversion (from simulation) are shown. The solid curves are the input turbulence strengths given, and the dashed curves are the results of the simulation. The two colors or markers represent two different sets of inputs and corresponding outputs. 
} 
\end{figure}

We tested for different turbulence conditions. Figure~\ref{fig:s_dimm_sim_res} shows two cases. The first case (blue curves with circular markers) has an ideal profile with turbulence decreasing with height. The second profile (orange curves with triangular markers) has one strong turbulence layer at 3000~m. For both cases, the input and output profiles are shown by solid and dashed curves, respectively. As stated in section \ref{sec:sim_procedure}, an ensemble average of 200 is done prior to fitting the theoretical covariance functions. The error bars in the output curves arise from averaging the results over multiple such sets of 200 matrices. The inversion is able to detect the given trends albeit not the exact input turbulence strengths. This trade-off is acceptable since we are interested in identifying the height of the strong layer of turbulence and not in measuring the exact strength of that layer. It can be seen that the strengths at 6000~m layer are consistently underestimated. The underestimation of the turbulence at higher heights is an intrinsic limitation of the S-DIMM+ method. While the code underestimates the turbulence from higher heights, it is still able to identify the strongest turbulence layer, which is the primary aim of this experiment. 



\section{Turbulence characterization with balloon-measurements}
\label{sec:balloon_theory}

\subsubsection{Principle of the method}

The temperature structure function ($D_T(r)$) is measured as the mean square temperature fluctuations between two points separated by a distance `$r$.’ By mounting two temperature sensors and having a mechanism to displace them in height, the temperature structure function can be measured as a function of height \cite{2004A&A...416.1193A, 2008PASP..120.1318M, 2009RaSc...44.2011R }. Then, assuming Kolmogorov's theory of turbulence, the temperature structure function parameter ($C_T^2$) can be found as \cite{1981PrOpt..19..281R}:

\begin{equation}
    \label{eq:dTtocT}
    D_T(r) = C_T^2r^{2/3}.
\end{equation}
Then, the turbulence strength profile, $C_N^2(h)$ can be estimated as:

\begin{equation}
    \label{eq:cTtocN}
    C_N^2(h) = \big(80*10^{-6} \frac{P(h)}{(T(h))^2} \big)^2C_T^2(h),
\end{equation}

where $P(h)$ is the pressure in milli bar and $T$ is the absolute temperature in Kelvin. Therefore, we have used a system (described in detail in \ref{sec:balloon_expt}) of seven Pt-100 sensors to determine the near-Earth turbulence, acting complementary to the S-DIMM+ method.

\section{Experimental setup and Analysis}
\label{sec:expt_obs}

The experiments were carried out at the Kodaikanal Observatory, Tamil Nadu, in January 2024. The S-DIMM+ experiment was conducted at the 38~cm KTT. A mask was used to reduce the telescope aperture to 36~cm. The balloon experiment was carried out near-simultaneously, about 20~m away from the telescope building. 

\subsection{S-DIMM+}

\subsubsection{Experimental Set-up}

\begin{figure}[H]
\begin{center}
\begin{tabular}{c}
\includegraphics[height=7cm]{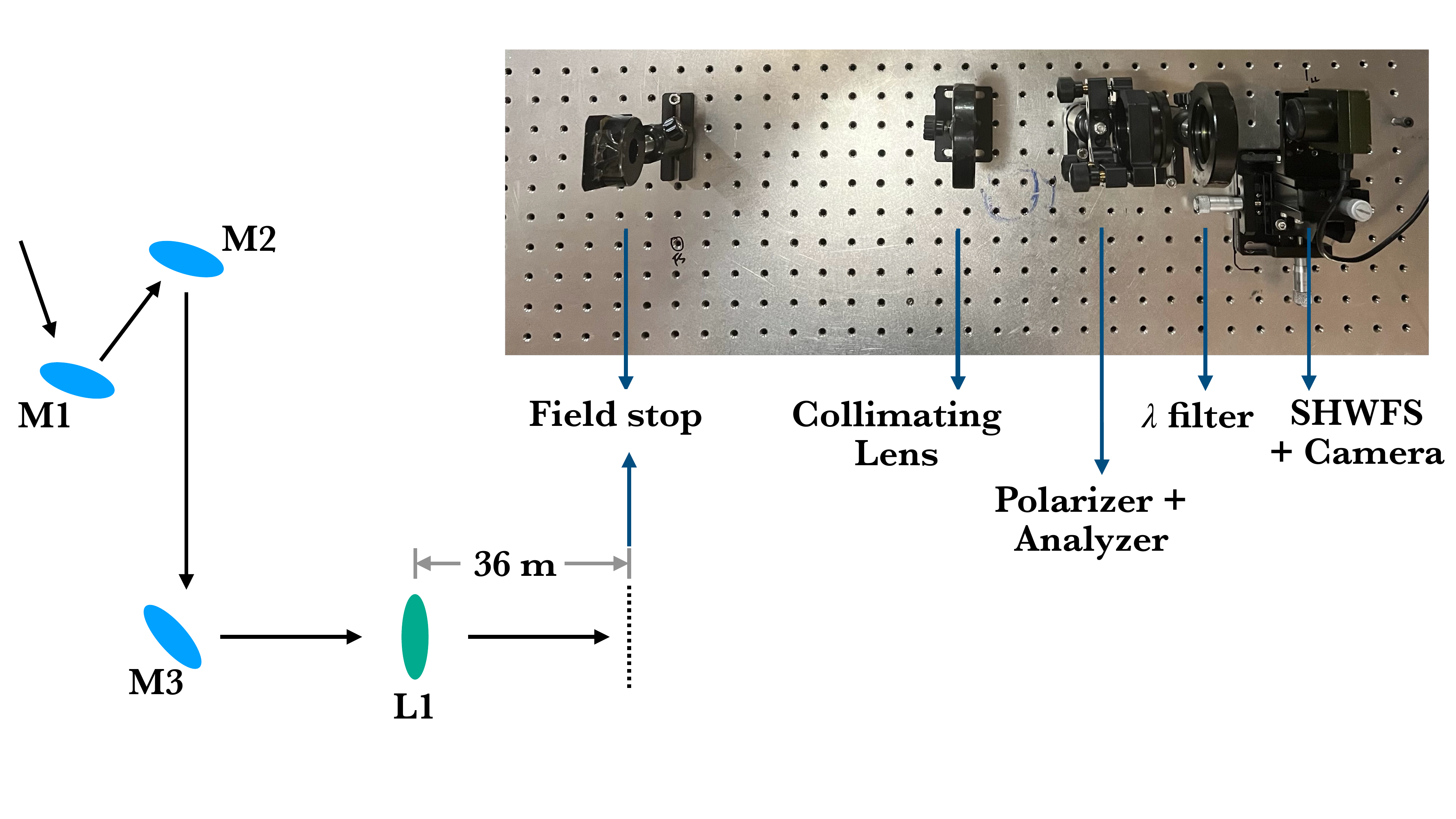}  \\
(a) \hspace{7cm} (b)
\end{tabular}
\end{center}
\caption 
{ \label{fig:S_DIMM_set_up_img} (a): Schematic of KTT (not to scale). The three mirrors (blue ovals) form the light feeding system, directing the light from the tower into a tunnel. The green oval is an achromatic doublet that produces an image of the sun at 36~m from it. (b): Optical setup at the telescope for S-DIMM+ measurements. 
} 
\end{figure}

Figure~\ref{fig:S_DIMM_set_up_img}~(a) shows the schematic of KTT. The first two mirrors form the coelostat (M1 and M2, indicated by two blue ovals) are on a tower roughly 11~m above the ground. M2 then directs the light vertically down onto M3, which reflects it into the tunnel. An achromatic doublet (green oval, L1) mounted on a rail then brings the light to a focus 36~m away from it (dotted line in schematic). At this focal plane, we used a field stop to select a region of about 23$\times$20 arc-seconds$^2$. Reimaging optics were used to reduce the beam diameter and fully illuminate eight lenslets across the diameter with an image scale of about 0.4$''$/pixel on the detector plane. Data was recorded using two wavelength filters - one at H-$\alpha$ (centered at 656.3~nm with 3.5~nm bandwidth) and one continuum filter (centered at 540~nm with 10~nm bandwidth). Additionally, a polariser-analyser combination was used to reduce the intensity of the incident light. Finally, an OKOTEK SHWFS (serial number: FS1540-H300-F18-16.04) mounted with a uEye camera (IDS UI-1540LE-M-GL66) was used to obtain the lenslet array images.

\begin{table}[H]
\caption{Observation set-up parameters} 
\label{tab:obs_param_setup}
\begin{center}       
\begin{tabular}{|l|l|} 
\hline
\rule[-1ex]{0pt}{3.5ex} Parameter & Value  \\
\hline\hline
 Telescope Primary Diameter & 36~cm \\
\hline
 Focal length & 36~m \\
\hline
  Wavelength(s) of observation & 656.3~nm and 540~nm \\
\hline
Focal length of collimating lens & 250~mm \\
\hline
 Pitch of SHWFS & 300 $\mu$m \\
\hline
 Pitch of SHWFS projected on pupil & 4.5~cm \\
\hline
Number of lenslets across Dia & 8\\
\hline
 Pixel size & 5.2 $\mu$m\\
\hline
 Full well capactiy & 40000 e$^-$\\
\hline
 Image scale & 0.413 $''$/pixel \\
\hline
Field-size of one sub-aperture image & 23$''\times$20$''$\\
\hline
Frame rate & 79 fps\\
\hline
Exposure time & 1~ms\\
\hline
\end{tabular}
\end{center}
\end{table}

\subsubsection{Data}
The measurements were done on the 16$^\text{th}$, 17$^\text{th}$ and 18$^\text{th}$ of January, 2024. We recorded bursts of data, each containing 2000 frames with an exposure time of about 1 ms. We have observed different active solar features (sunspots or pores) for our analysis. 

\begin{figure}[H]
\begin{center}
\begin{tabular}{c}
\includegraphics[height=8cm]{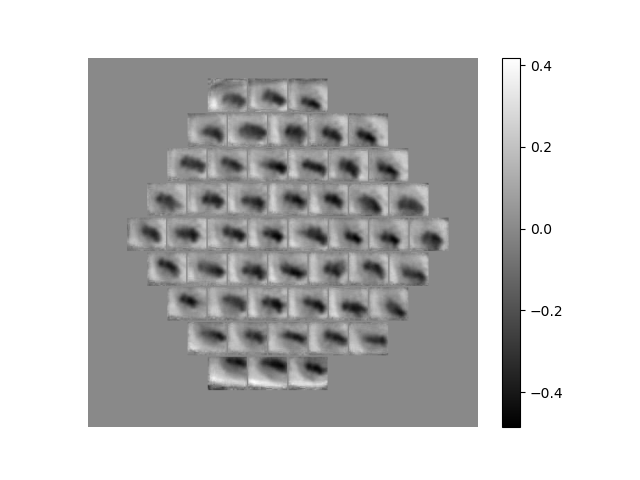}
\end{tabular}
\end{center}
\caption 
{ \label{fig:sun_spot_imgs} One frame containing the images of a sunspot as recorded by lenslet array at 540~nm (exposure time $\sim$ 1 ms). Each image has a field of view of about 23$\times$20 arc-second$^2$.
Standard image processing steps of dark and flat correction have been done. The images are also made zero mean. 
} 
\end{figure}

\subsubsection{Data Analysis}

The procedure was as follows:
\begin{itemize}
    \item Each recorded image frame contains fifty sub-images corresponding to the fifty fully illuminated lenslets. Out of these, only the eight across the diameter, along with one other image (identified as the one with the highest contrast, called reference image), are used for the analysis. 
    \item The regions where these nine images lie are identified. Each subimage occupies 56x48 pixels. The corresponding regions of the flat and dark images are used for calibration.
    \item The covariance matrices are estimated as described in section \ref{sec:SDIMMplusprinciple} using two subaperture images and two subfields within them at a time. Instead of the absolute image motion, the differential image motion is found using the reference image.
    \item 200 covariance matrices were ensemble averaged and then fitted to the theoretical equations to estimate the strength of turbulence. 
\end{itemize}

\subsubsection{Results - High-altitude turbulence from S-DIMM+} 
\label{sec:S_DIMM_res}

\begin{figure}[H]
\begin{center}
\begin{tabular}{c}
\includegraphics[trim = 0 6cm 0 0, clip = True, height = 8cm]{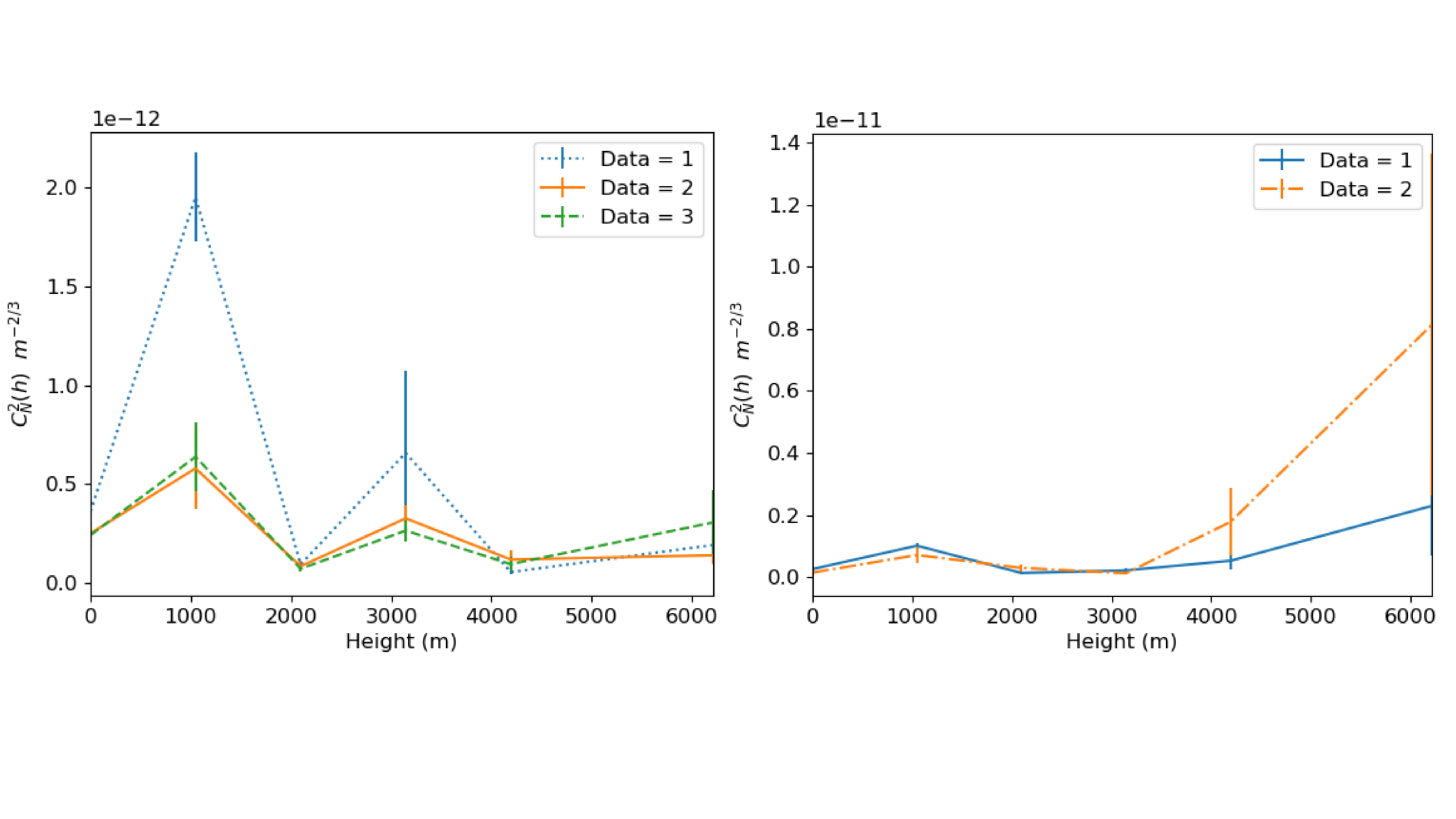} \\
(a) \hspace{7cm} (b)
\end{tabular}
\end{center}
\caption 
{ \label{fig:S_DIMM_res} Results from the S-DIMM+ inversion. (a) Results from three sets of data taken before zenith. (b) Results from two sets of data taken after zenith. The profiles are not airmass corrected.
} 
\end{figure}

Figure~\ref{fig:S_DIMM_res}~(a) and (b) show the results obtained from the S-DIMM+ method pre and post-local noon, respectively. In the former, the dotted blue and solid orange curves (Data 1 and 2) are obtained from data recorded at 656.3~nm, and the dashed green curve using data recorded at 540~nm (Data 3). Data set 1 was observed on 16$^\text{th}$ January, 2024 at about 6:45 UT. Data set 2 and 3 were recorded on the 18$^\text{th}$ of January, 2024 at about 6:45 UT and 5:30 UT, respectively.

Similarly, in the post-noon curves, the solid blue curve was estimated with data recorded at 656.3~nm, and the dashed orange one used data taken at 540~nm. The former data set was observed on the 17$^\text{th}$ of January, 2024 at about 8:50 UT and the latter on 18$^\text{th}$ of January, 2024 at about 7:15 UT. In both the pre- and post-noon cases, each data set corresponds to four bursts of 2000 frames each. The time gap between two bursts (within one data set) was about 3 - 5~min. The error bars arise from averaging over the inversion results from the four bursts of data in each data set. From the figures, it can be seen that our inversion code is not wavelength-dependent.

Before noon, we identified the presence of a strong turbulence layer at about 3~km above the telescope. Another strong layer at roughly 1~km above the telescope can also be seen. But, as stated in section \ref{sec:inv_code}, our code was not sensitive to turbulence at this height. Therefore, the detected turbulence at 1~km may be due to a systematic from the code or due to an actual strong layer at that height. We also saw the evolution of turbulence during the day, with the seeing worsening with time. This is expected for a mountain site like Kodaikanal, as the ground heats up as the day progresses. We could observe this effect visually as the degradation in the quality of the SHWFS images. This can also be seen in Figure~\ref{fig:S_DIMM_res}~(b), which shows that the turbulence has increased by an order of magnitude compared to the pre-noon profile. The error bars associated with this profile are also higher as the inversion is known to fail under poor seeing conditions \cite{2010A&A...513A..25S}.

\subsection{Balloon-Measurements}
\label{sec:balloon_expt}

\subsubsection{Experimental set-up}

\begin{figure}[H]
\begin{center}
\begin{tabular}{c}
\includegraphics[height=6cm]{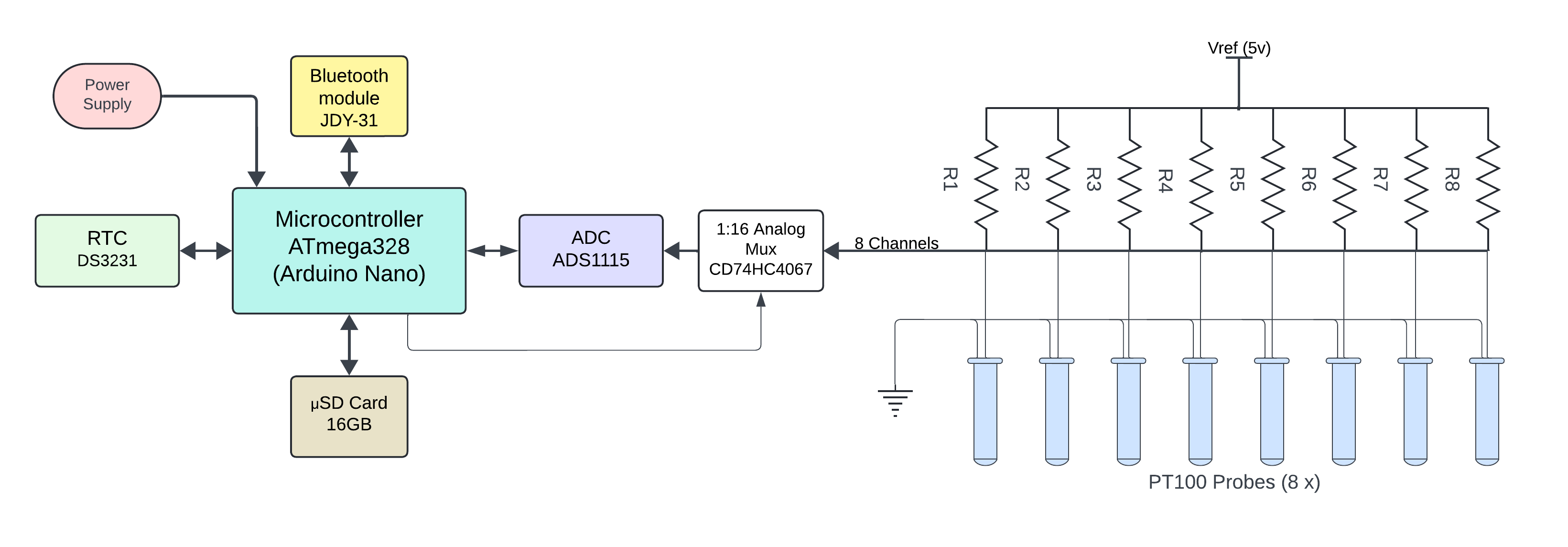}
\end{tabular}
\end{center}
\caption 
{ \label{fig:block_diag_DL} Block diagram of the electronic components of IDL. 
} 
\end{figure}

We have used latex balloons filled with hydrogen. The balloons were procured from Pawan Balloon (Pune, India). The hydrogen gas (99.99\%) was procured from Sri Venkateshwara Carbonic Gases Pvt Ltd (Coimbatore, India). The volume of hydrogen gas required for a payload of about 3.2 kg was 4.7 m$^3$ \cite{2013arXiv1302.0981N, 2016A&AT...29..397S}. The payload consists of two parts, an IP65 waterproof plastic enclosure (Figure~\ref{fig:balloon_data_log_pics}~(b)) housing all the electronics components and a rod mounted with seven immersion-type Pt-100 sensors.
The IP65 enclosure was further kept in a Styrofoam box during the balloon flight. The electronics setup involves a custom-built microcontroller-based data logger called IIA Data Logger (IDL). At its heart is the ATmega 328 Arduino Nano microcontroller, which receives sensor readings from an Analog to Digital Converter (ADC - ADS1115) and timing information from a Real Time Clock (RTC - DS3231). It saves the collected sensor data along with the time stamp onto the SD Card. A set of eight immersion-type Pt-100 temperature sensors connected in resistance divider mode with eight precision resistances (5K Ohm) are used. Only seven of these sensors were used during the actual experiment as one was kept as backup to be used in case one of the other sensors failed. The voltage across the Pt-100 is measured using 16bit ADC (ADS1115) with an input range configured to +/-0.256 V using an internal Programmable Gain Amplifier (PGA). All the sensors are sampled using the same ADC one after the other. The sensor to be sampled is selected by using an analog multiplexer (CD74HC4067) controlled by the microcontroller. The various components and their interactions are shown as a block diagram in Figure~\ref{fig:block_diag_DL}.

For calibration,  resistance with a known value was connected in the place of the Pt-100 sensor, and the ADC counts were recorded and mapped to a Pt-100 temperature vs resistance table as described in Appendix~A. Each measurement happens in a burst mode every 1 sec, with 10 sets of measurements of all the 8 sensors. The time required for sampling data from all eight sensors is 80 ms, with 10 ms between each sensor reading. Since this is a field experiment, provision of an LED indicator is given to indicate data is being recorded. Additionally, Bluetooth connection allows users to connect with the microcontroller and receive present data with a time stamp for sensor functionality check. The entire system works with a 7.4 V, 2200 mAh Lithium Polymer battery pack and can last up to 24 hours of operation.

\begin{figure}[H]
\begin{center}
\begin{tabular}{c}
\includegraphics[height=7cm]{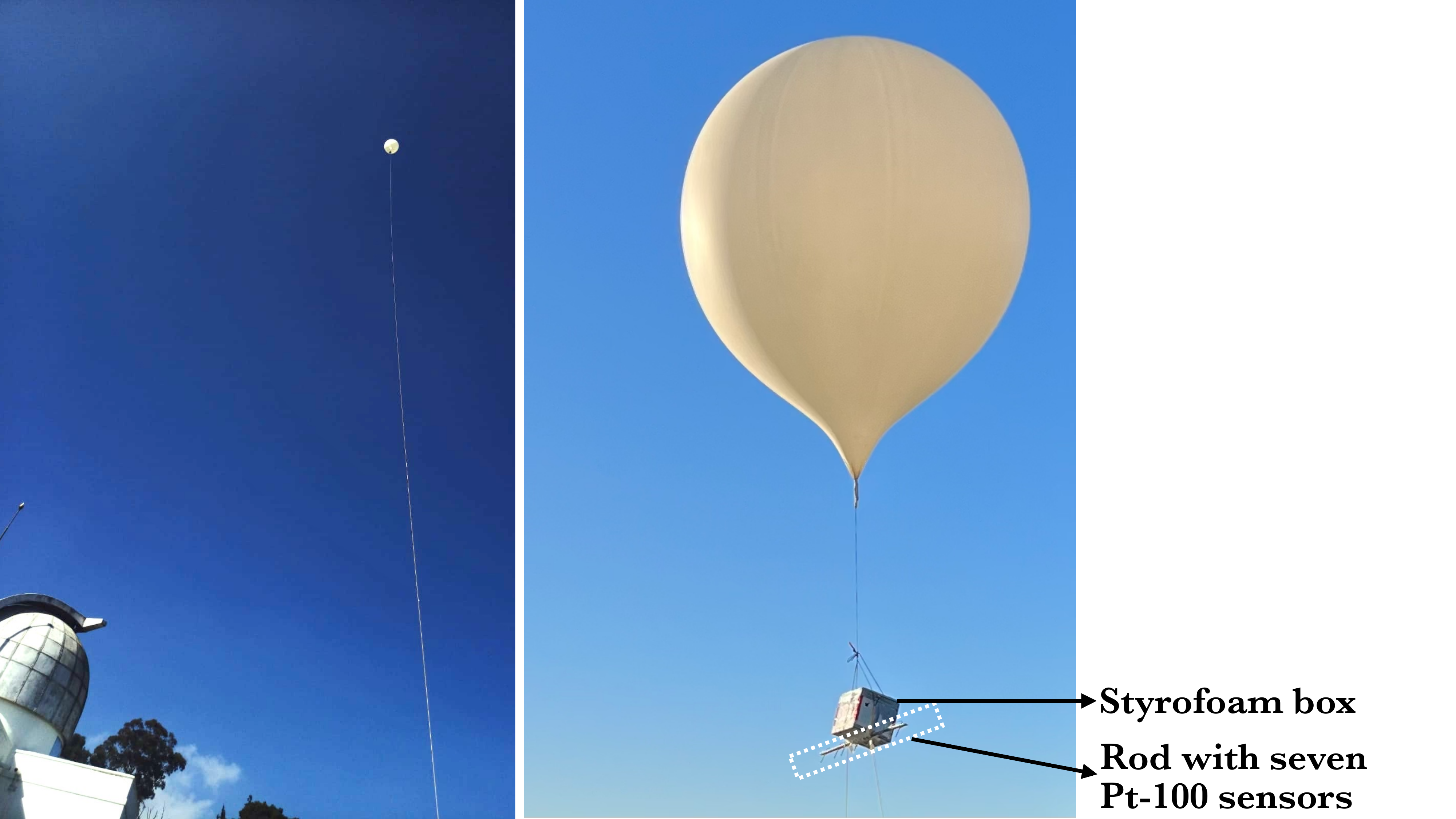}  \\
(a) \hspace{4cm} (b)
\end{tabular}
\end{center}
\caption 
{ \label{fig:balloon_kodai} 
Photographs of the tethered balloon during measurements. (a) Balloon-measurements setup with the dome of KTT on the bottom left. (b) Zoomed picture of the setup. A white dotted box marks the rod with Pt-100 sensors mounted. A thermally insulated Styrofoam box houses the electronics.} 
\end{figure}

\subsubsection{Data and analysis}
For the profile measurement experiments, the seven Pt-100 sensors were mounted on a flat rod about 1.3~m long. This allows 21 possible combinations of baselines when the sensors are chosen two at a time. There were a total of eleven unique baselines from 10~cm to 110~cm in steps of 10~cm. Two holes were made at the ends of the rod to allow a Nylon rope to tie it to the balloon. Additionally, another rope with markings every 10~m was used to hold and hoist the balloon. It was kept at a given height for one minute and then moved to the next height (separated by 10~m). There is roughly a 10~m offset between the ``ground level'' of the balloon-measurements and that of S-DIMM+ (GL-(S-DIMM+)). The duration to move between heights was about 20 s. Then, an additional 1 min 40 s was allowed for the medium to settle down. Therefore, for each height, around 600 points of data were recorded by each of the seven sensors. There were some missing points.  

The analysis procedure was as follows:
\begin{itemize}
    \item For each height, the counts recorded by each sensor were collected.
    \item Using the calibration method described in Appendix~A, the temperature recorded by each sensor was calculated. 
    \item The bias offset of each sensor was removed.
    \item Using two sensors at a time, the temperature structure function was found as the mean square difference of the temperature fluctuations between the points. 
    \item $C_T^2$  and then $C_N^2$ were found using equations \ref{eq:dTtocT} and \ref{eq:cTtocN} respectively. By using equation \ref{eq:dTtocT}, we are assuming a Kolmogorov model of atmospheric turbulence.
    For substituting in \ref{eq:cTtocN}, the pressure (in millibar) at a given height ($h$) above the ground can be expressed as \cite{press_eqn}:
\begin{equation}
    \label{equ:press}
    P(h) = \left( \frac{44331.514 - h}{11880.516}\right)^\frac{1}{0.1902632} .
\end{equation}

\end{itemize}

The above process is repeated for the twenty-one possible combinations of two sensors. The $C_N^2$ profiles obtained from each are averaged to get the cumulative profile. An important point is that there is a finite time delay between the measurements made by the different sensors. The delay is a function of the two sensors chosen. This is a consequence of opting for a simpler setup to ease the calibration procedure by using a single ADC for all the sensors.

\subsubsection{Results - Near-Earth turbulence}
Three data sets were recorded on the 17$^\text{th}$ of January 2024. 
The first set started at about 5:10 UT (10:40 AM local time) and lasted about an hour. The balloon’s initial height was 10~m, and its final height was 180~m. The second set started immediately after the first set at about 6:00 UT. This also lasted for about an hour, and data was recorded from 180~m to 10~m (in the descent phase). The third set started around 8:45 UT (after local meridian transit) and lasted almost two hours. 

\begin{figure}[H]
\begin{center}
\begin{tabular}{c}
\includegraphics[trim = 0 6cm 0 0, clip = True, height = 8cm]{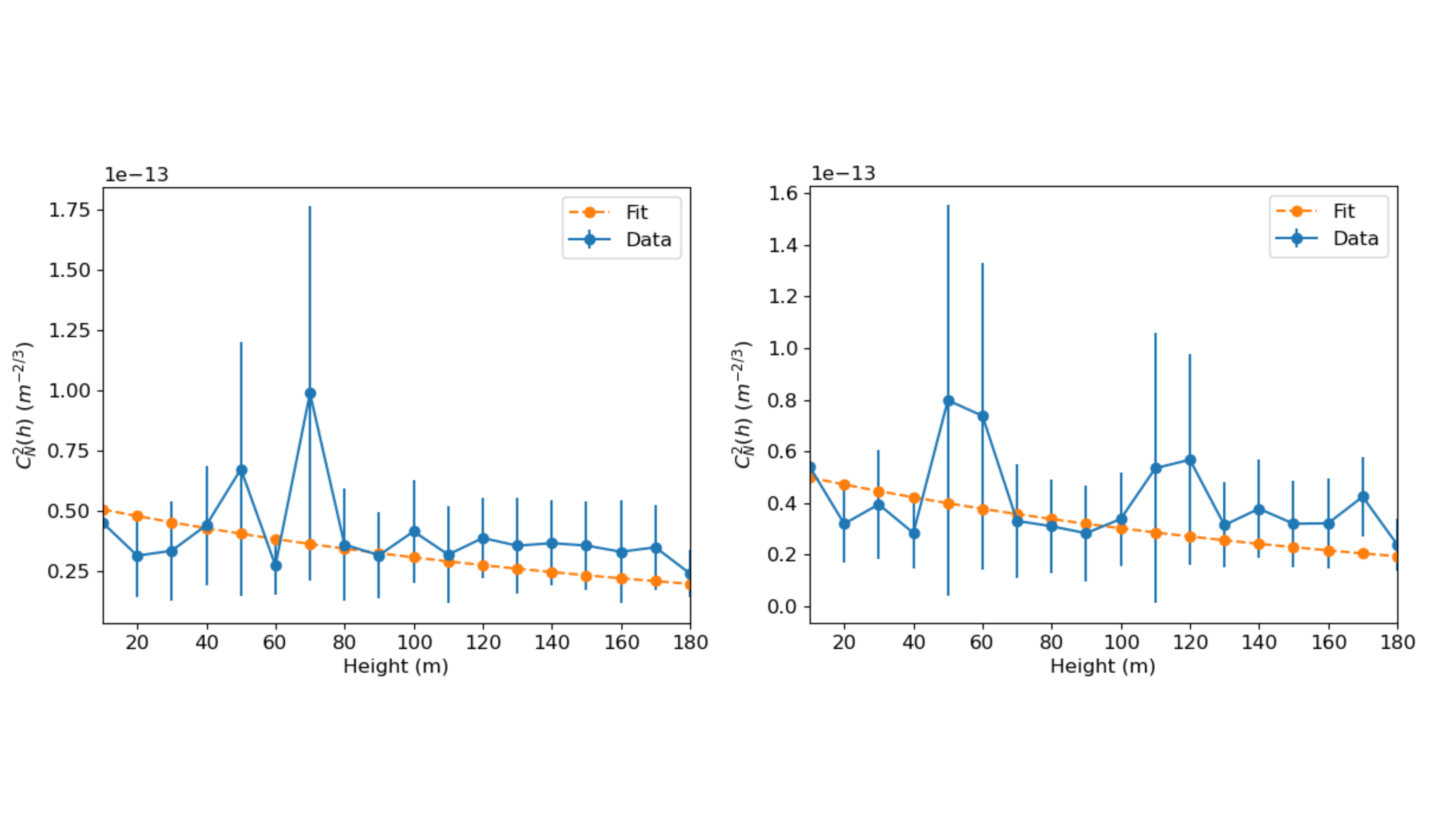}  \\
(a) \hspace{7cm} (b)
\end{tabular}
\end{center}
\caption 
{ \label{fig:balloon_res_AM} Results from the balloon experiment taken on 17$^\text{th}$ before local meridian transit. (a) From data set 1 (ascent phase) and (b) set 2 (descent phase). The solid blue curves in the two plots are from the values calculated from the experiment, and the dashed orange curves are from the fit of the daytime component. For the fits shown here, $\frac{V_w}{\overline{V_w}}$ = 3 and $h_0$ = 180~m were used.
} 
\end{figure} 

\begin{figure}[H]
\begin{center}
\begin{tabular}{c}
\includegraphics[height=6cm]{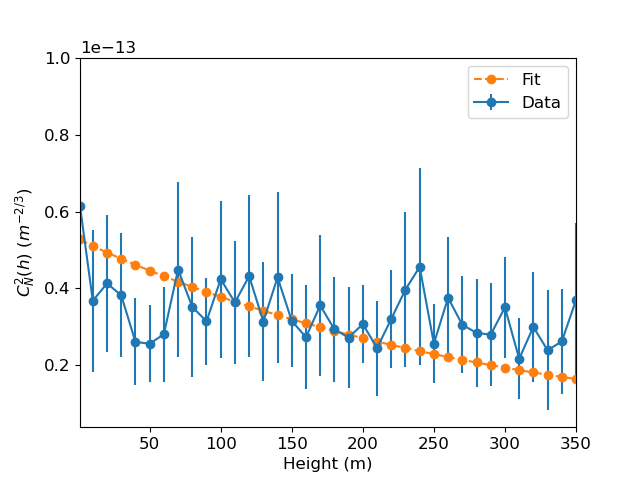}
\end{tabular}
\end{center}
\caption 
{ \label{fig:balloon_res_PM} Results from the balloon experiment taken on 17$^\text{th}$ after local meridian transit. The color (or style) of the curves are the same as those in Figure~\ref{fig:balloon_res_AM}. For the fit shown here,  $\frac{V_w}{\overline{V_w}}$ = 3 and $h_0$ = 300.
} 
\end{figure}

The balloon reached a maximum height of 350~m. These three sets will hereafter be referred to as sets 1, 2, and 3, respectively. The three measured $C_N^2$ profiles are shown by solid blue curves in Figure~\ref{fig:balloon_res_AM}~(a), (b), and Figure~\ref{fig:balloon_res_PM} respectively.

The cumulative $C_N^2$ profiles of each data set are fitted to the modified Hufnagel model. The Hufnagel model is given by \cite{hufnagel}: 
\begin{equation}
    \label{eq:huf_vall}
    C_{N_{\textit{HV}}}^2(h) = A\left[  2.2\times 10^{-23} \left( \frac{z+h}{1000} \right) ^{10} exp \left( \frac{-z+h}{1000}\right) \left( \frac{V_w}{\overline{V_w}}\right)^2 + 10^{-16}exp\left(\frac{-z+h}{1500}\right)   \right]   ,
\end{equation}
where `$A$' is a scaling constant, `$h$' is the height above the ground in m, `$z$' is the elevation of the site in m, and $\frac{V_w}{\overline{V_w}}$ is the ratio of upper to mean atmospheric wind speeds. 
An additional term must be added to account for strong daytime turbulence, and the total profile is given by \cite{atst_site_survey}:
\begin{equation}
    \label{eq:dat_time_cn2}
    C_N^2(h) = C_{N_{\textit{HV}}}^2(h) + A_Bexp\left(\frac{-h}{h_0}\right) ,
\end{equation}

where $A_B$ is the boundary amplitude and $h_0$ is the boundary scale height.

We initially fitted the three data sets for the model given by Equations \ref{eq:huf_vall} and \ref{eq:dat_time_cn2}. Then, since the balloon has a maximum measurement height of 350 m, we fitted only for the daytime component described by the second term of Equation \ref{eq:dat_time_cn2}. In both cases, we assumed different wind speed ratios $\left( \frac{V_w}{\overline{V_w}} \right)$ and boundary scale heights ($h_0$). We found that fitting only for the daytime component did not significantly change the results when compared to the fitting for the full model. This shows that the daytime component dominates in near-Earth turbulence. The term $A_B$ was used as a parameter for the fit. The integral of the $C_N^2$ profile also gives the integrated $r_0$.

The results of the fit are shown graphically in Figures~\ref{fig:balloon_res_AM}~(a), (b) and Figure~\ref{fig:balloon_res_PM} as dashed orange curves. It can be seen from these figures that the experiment is able to detect the overall trend predicted by the daytime component. Once the parameter $A_B$ was estimated for each curve, it was used to generate a model atmosphere at finer sampling (every 20~cm). The $r_0$ is estimated with the $C_N^2$ profile modeled using a finer sampling. An earlier study\cite{kodai_iiap} at the observatory used image motions to determine the seeing. The median $r_0$ was estimated to be about 3.9~cm at 500~nm. We find that our $r_0$ estimates, when measured from a fine sampled $C_N^2$ profile, match the earlier results well. These results for the three data sets are shown in Tables~\ref{tab:balloon_fit}, \ref{tab:balloon_fit2}, and \ref{tab:balloon_fit3}, respectively. From the tables, it can be seen that for a given data set, varying the wind velocity ratio and boundary scale height does not significantly change the results of the fit. However, we did note that a boundary scale height ($h_0$) of 300~m gave a better fit for data set 3 compared to the other two sets having a good fit for a $h_0$ of 180~m. We need to repeat the experiment to understand if this is due to the diurnal variation. Furthermore, from Figures~\ref{fig:S_DIMM_res}~(a),  \ref{fig:balloon_res_AM} and \ref{fig:balloon_res_PM}, we can see that the order of magnitude of the $C_N^2$ profile values from the two experiments also match.

\begin{table}[H]
\caption{Results from balloon-measurements for data set 1. The measured data was fitted to the daytime contribution model (Equation \ref{eq:dat_time_cn2} second term). The results of the fit for different combinations of wind velocity ratios and boundary scale height (first and second columns, respectively) are shown below. The fitted values of $A_B$ are shown in the third column. The $r_0$ (at 500~nm) estimated from the fit at finer sampling are given in the last column. $r_0$ estimated using measured profile from data set~1 was 38.79~cm (at 500~nm). } 
\label{tab:balloon_fit}
\begin{center}       
\begin{tabular}{|c|c|c|c|} 
\hline
\rule[-1ex]{0pt}{3.5ex} $\frac{V_w}{\overline{V_w}}$ & $h_0$ & $A_B$  & $r_0$ (cm) - from fit \\
& (m) &    & 20~cm sampling\\
\hline\hline
 0.3 & 100  & 6.47e-14 & 4.76 \\
 3   & 100  & 6.47e-14 & 4.76 \\
 0.3 & 180  & 5.34e-14 & 4.35 \\
 3    & 180 & 5.33e-14 & 4.35 \\
\hline
\end{tabular}
\end{center}
\end{table}

\begin{table}[H]
\caption{Results from balloon-measurements for data set 2. The column headers are the same as \ref{tab:balloon_fit}. $r_0$ estimated using measured profile from data set~2 was 38.21~cm (at 500~nm).} 
\label{tab:balloon_fit2}
\begin{center}       
\begin{tabular}{|c|c|c|c|} 
\hline
\rule[-1ex]{0pt}{3.5ex} $\frac{V_w}{\overline{V_w}}$ & $h_0$ & $A_B$  & $r_0$ (cm) - from fit \\
& (m) &    & 20~cm sampling\\
\hline\hline
 0.3 & 100  & 6.42e-14 & 4.85 \\
 3   & 100  & 6.42e-14 & 4.85 \\
 0.3 & 180  & 5.27e-14 & 4.48 \\
 3    & 180 & 5.27e-14 & 4.48 \\
\hline
\end{tabular}
\end{center}
\end{table}

\begin{table}[H]
\caption{Results from balloon-measurements for data set 3. The column headers are the same as \ref{tab:balloon_fit}. $r_0$ estimated using measured profile from data set~3 was 28.49~cm (at 500~nm).} 
\label{tab:balloon_fit3}
\begin{center}       
\begin{tabular}{|c|c|c|c|} 
\hline
\rule[-1ex]{0pt}{3.5ex} $\frac{V_w}{\overline{V_w}}$ & $h_0$ & $A_B$  & $r_0$ (cm) - from fit \\
& (m) &    & 20~cm sampling\\
\hline\hline
 0.3 & 180  & 6.12e-17 & 3.17 \\
 3   & 180  & 6.12e-17 & 3.17 \\
 0.3 & 300  & 5.28e-14 & 2.91 \\
 3    & 300 & 5.28e-14 & 2.91 \\
\hline
\end{tabular}
\end{center}
\end{table}

\section{Conclusions}
\label{sec:result}

We have used the S-DIMM+ method and a balloon-borne array of temperature sensors to measure the higher-altitude and near-Earth turbulence at Kodaikanal Observatory. While validating the performance of the former method for our system parameters through simulations, we have tested the primary assumptions in the principle of this method. We have also experimentally established that even a single small telescope can be used to measure the turbulence strength up to a height of 5 - 6~km. This helped us identify a strong layer of turbulence about 3~km above the telescope site based on the preliminary experiments. Further measurements can be carried out to understand the evolution of turbulence as a function of diurnal and seasonal variations. A larger dataset will also help us achieve a statistically significant result.
 
The balloon-measurements for near-Earth turbulence also proved successful and have a reasonable match to the profile predicted by Hufnagel (modified for daytime). One problem we encountered with the balloon was the slating of the balloon due to wind. We have ignored the effect of this in the present analysis. This causes an error in the height measurement. If a drone is used in the place of the balloon, this effect can be alleviated. There is also a temporal delay between the measurements of the different temperature sensors. Ideally, all the sensors must be read at the same instant of time. One possible way of overcoming this would be to use a Wheatstone bridge-like setup and measure the temperature difference between the two sensors directly \cite{2005PASP..117..536A}. Sensors with a faster response time will also further improve the accuracy of the measurements.

In conclusion, we have used two different methods to estimate the $C_N^2$ profile for the first time at Kodaikanal Observatory. These experiments can be repeated at any other site to understand the turbulence strength profile there. The S-DIMM+ method will be particularly useful in identifying strong turbulence layers for MCAO. The balloon-measurements can also be used to identify the height of future ground-based solar telescopes like the National Large Solar Telescope (NLST)\cite{2008JApA...29..345S}.

\appendix    

\section{Calibration of Pt-100 sensors}

\begin{figure}[H]
\begin{center}
\begin{tabular}{c}
\includegraphics[height=5.5cm]{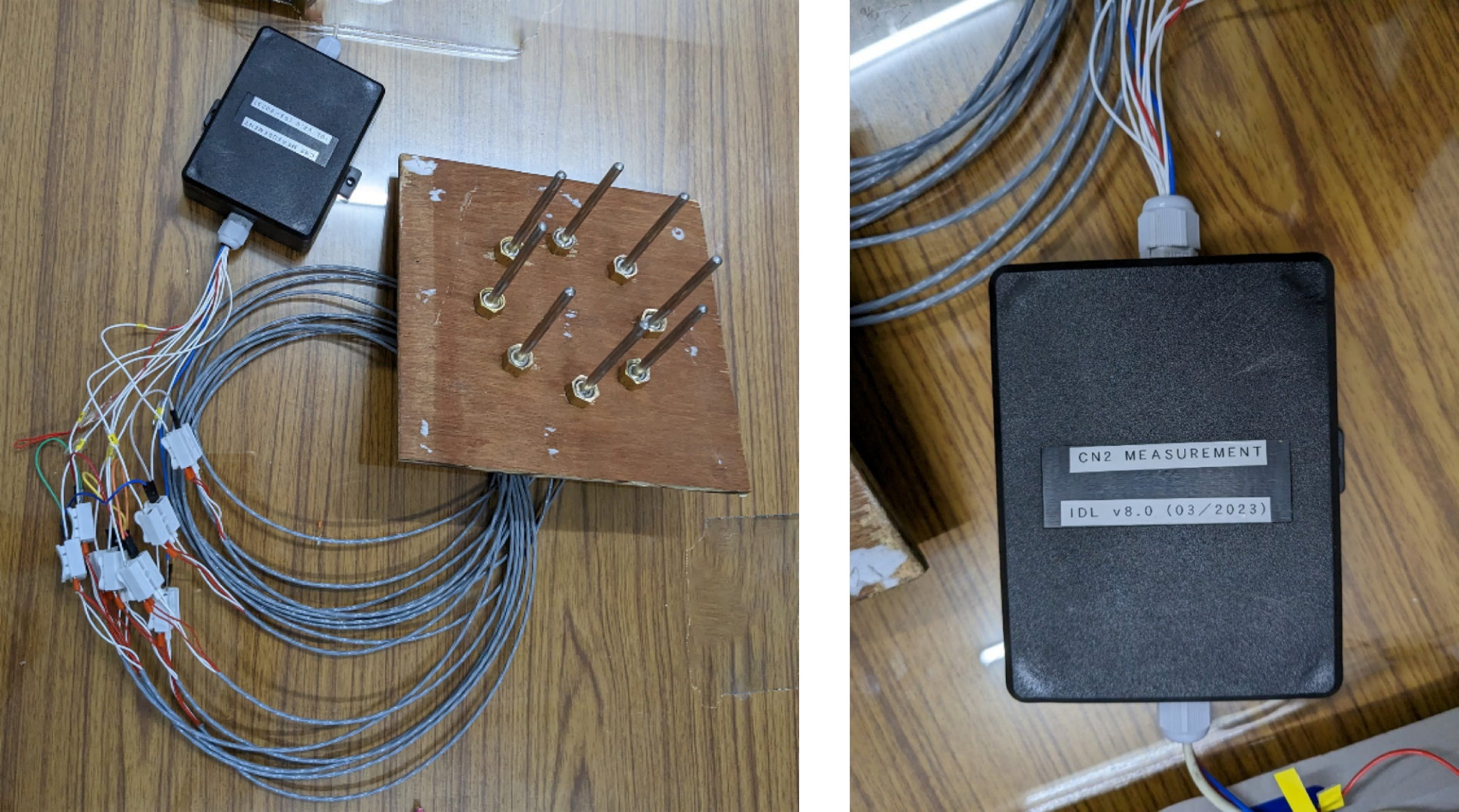} \\
(a) \hspace{6cm} (b)
\end{tabular}
\end{center}
\caption 
{ \label{fig:balloon_data_log_pics} 
Pictures of the balloon-measurements set-up taken during calibration. (a) The immersion-type Pt-100 temperature sensors were mounted on a wooden block in a circular fashion for convenience during calibration when they were immersed in ice/boiling water. (b) Photograph of IP65 enclosure enclosing the electronics used for this experiment.}
\end{figure} 

\begin{figure}[H]
\begin{center}
\begin{tabular}{c}
\includegraphics[height=10cm]{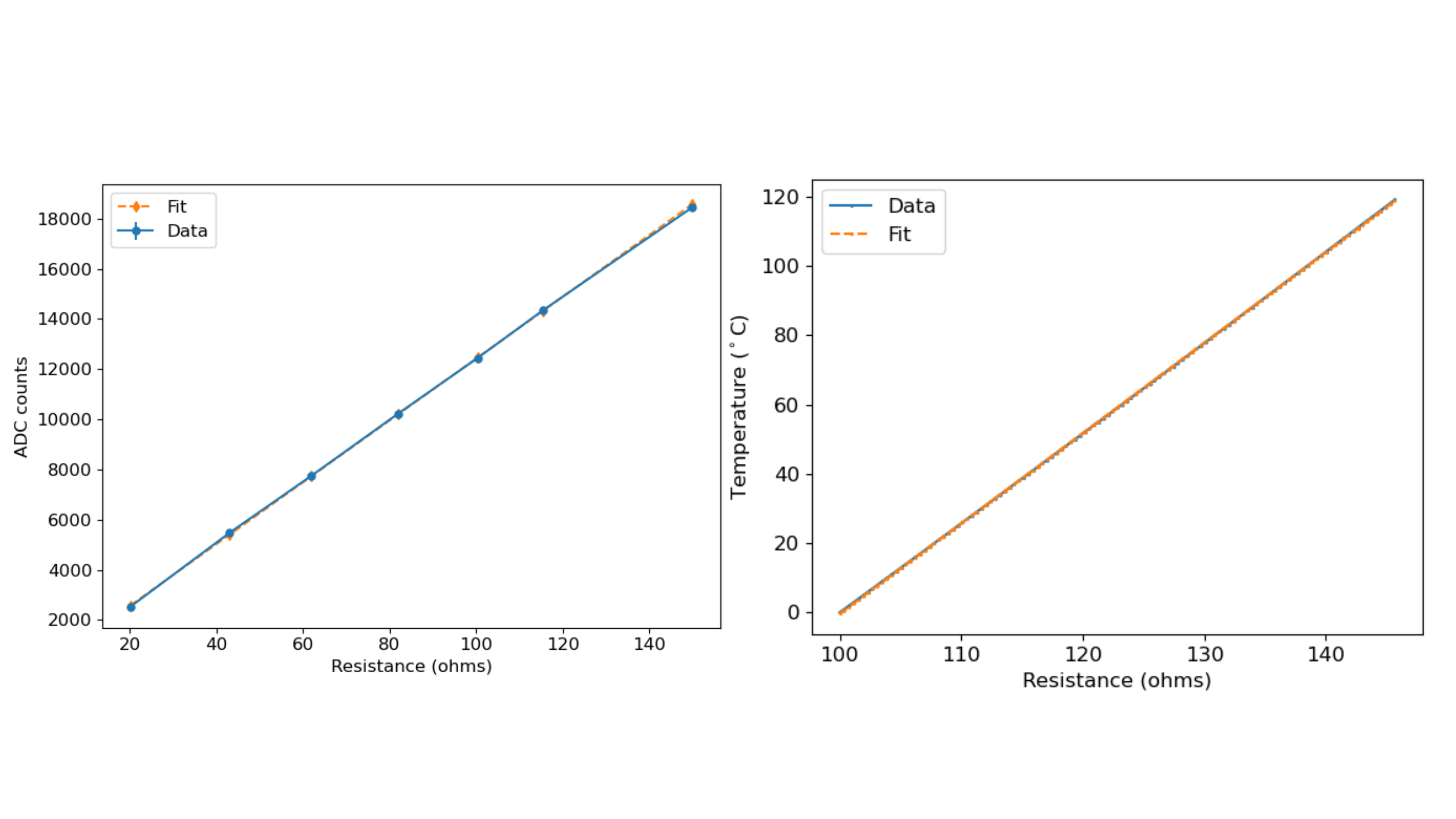} \\
(a) \hspace{6cm} (b)
\end{tabular}
\end{center}
\caption 
{ \label{fig:adc_vs_resis} 
Calibration curves for balloon-measurements. (a) ADC Counts vs Resistance. (b) Temperature vs Resistance.} 
\end{figure}

The digital counts recorded by the Pt-100 sensors need to be calibrated to get the actual temperature values. The photographs shown in Figure~\ref{fig:balloon_data_log_pics} were taken during the calibration, which was done in two stages. First, a resistance with a known value was connected to the circuit in the place of the Pt-100 sensor, and the counts were recorded. This was repeated with different resistances. A straight line was fit, and the values convert counts to resistances. This is shown in Figure~\ref{fig:adc_vs_resis}~(a). The solid blue curve is from the data recorded, and the dashed orange curve is the straight line fit. The error bars of the former curve arise from averaging the data counts recorded over a few seconds. The magnitude of the error bars is much smaller than the actual values of the counts.

Then, the resistances were converted to temperature values using the standard table (European Standard of Pt-100 Temperature vs Resistance Table). We also verified if the Pt-100 performance was as per the specification by measuring the sensor’s resistance at the ice point and boiling point of water. This is shown in Figure~\ref{fig:adc_vs_resis}~(b).

\subsection*{Disclosures}
The authors declare no conflicts of interest. 

\subsection* {Code, Data, and Materials Availability} 
The authors received the simulated solar granulation images from another source (Prof. S. P. Rajaguru of the Indian Institute of Astrophysics) and, therefore cannot share the same. The codes developed for the simulation and analysis of the data are not publicly available at this time.

\subsection* {Acknowledgments}
We thank Prof. S. P. Rajaguru for providing us with simulated solar granulation images, which were used to develop and test the inversion code. We thank Mr. Devendran, Mr. Kumaravel, and Mr. Ramesh of Kodaikanal Observatory for their assistance with the KTT experiments. We would also like to thank the staff of Kodaikanal Observatory (KO) for their help with the balloon experiment. We would like to thank the referees for their comments and suggestions, which have helped improve the paper.
This research has made use of NASA’s Astrophysics Data System Bibliographic Services.


\bibliography{main}   

\begin{thebibliography}{10}

\bibitem{1953PASP...65..229B}
H.~W. {Babcock}, ``{The Possibility of Compensating Astronomical Seeing},'' {\em PASP} {\bf 65}, 229  (1953).

\bibitem{1986ESOC...24..229S}
M.~{Sarazin}, ``{ESO VLT site evaluation. II.},'' in {\em European Southern Observatory Conference and Workshop Proceedings},  S.~{D'Odorico} and J.~P. {Swings}, Eds., {\em European Southern Observatory Conference and Workshop Proceedings} {\bf 24}, 229--238  (1986).

\bibitem{1988SoPh..115..183Z}
H.~{Zirin} and J.~M. {Mosher}, ``{The Caltech solar site survey, 1965 1967},'' {\em Solar Phys.} {\bf 115}, 183--202  (1988).

\bibitem{2005PASP..117.1296S}
H.~{Socas-Navarro}, J.~{Beckers}, P.~{Brandt}, {\em et~al.}, ``{Solar Site Survey for the Advanced Technology Solar Telescope. I. Analysis of the Seeing Data},'' {\em PASP} {\bf 117}, 1296--1305  (2005).

\bibitem{2016SPIE.9909E..6XM}
J.~{Marco de la Rosa}, L.~{Montoya}, M.~{Collados}, {\em et~al.}, ``{Daytime turbulence profiling for EST and its impact in the solar MCAO system design},'' in {\em Adaptive Optics Systems V},  E.~{Marchetti}, L.~M. {Close}, and J.-P. {V{\'e}ran}, Eds., {\em Society of Photo-Optical Instrumentation Engineers (SPIE) Conference Series} {\bf 9909}, 99096X  (2016).

\bibitem{1985ARA&A..23...19C}
C.~E. {Coulman}, ``{Fundamental and applied aspects of astronomical ``seeing''.},'' {\em Annu Rev. Astron. Astrophys.} {\bf 23}, 19--57  (1985).

\bibitem{1972JOSA...62.1068B}
J.~L. {Bufton}, P.~O. {Minott}, M.~W. {Fitzmaurice}, {\em et~al.}, ``{Measurements of Turbulence Profiles in the Troposphere},'' {\em Journal of the Optical Society of America (1917-1983)} {\bf 62}, 1068  (1972).

\bibitem{1976JOSA...66.1380B}
R.~{Barletti}, G.~{Ceppatelli}, L.~{Paterno}, {\em et~al.}, ``{Mean vertical profile of atmospheric turbulence relevant for astronomical seeing},'' {\em Journal of the Optical Society of America (1917-1983)} {\bf 66}, 1380  (1976).

\bibitem{1990A&A...227..294S}
M.~{Sarazin} and F.~{Roddier}, ``{The ESO differential image motion monitor},'' {\em A\&A} {\bf 227}, 294--300  (1990).

\bibitem{1987PASP...99.1360M}
H.~M. {Martin}, ``{Image motion as a measure of seeing quality},'' {\em PASP} {\bf 99}, 1360--1370  (1987).

\bibitem{2002PASP..114.1156T}
A.~{Tokovinin}, ``{From Differential Image Motion to Seeing},'' {\em PASP} {\bf 114}, 1156--1166  (2002).

\bibitem{2001ExA....12....1B}
J.~M. {Beckers}, ``{A Seeing Monitor for Solar and Other Extended Object Observations},'' {\em Experimental Astronomy} {\bf 12}, 1--20  (2001).

\bibitem{2011MNRAS.416.2154K}
T.~{Kawate}, Y.~{Hanaoka}, K.~{Ichimoto}, {\em et~al.}, ``{Seeing measurements using the solar limb - I. Comparison of evaluation methods for the Differential Image Motion Monitor},'' {\em MNRAS} {\bf 416}, 2154--2162  (2011).

\bibitem{1981PrOpt..19..281R}
F.~{Roddier}, ``{The effects of atmospheric turbulence in optical astronomy},'' {\em Progess in Optics} {\bf 19}, 281--376  (1981).

\bibitem{1993SoPh..145..389S}
E.~J. {Seykora}, ``{Solar Scintillation and the Monitoring of Solar Seeing},'' {\em Solar Phys.} {\bf 145}, 389--397  (1993).

\bibitem{1993SoPh..145..399B}
J.~M. {Beckers}, ``{On the Relation Between Scintillation and Seeing Observations of Extended Objects},'' {\em Solar Phys.} {\bf 145}, 399--402  (1993).

\bibitem{1999ASPC..184..309B}
J.~M. {Beckers}, ``{The Determination of Seeing, Isoplanatic Patch Size and Coherence Time by Solar Shadow Band Ranging},'' in {\em Third Advances in Solar Physics Euroconference: Magnetic Fields and Oscillations},  B.~{Schmieder}, A.~{Hofmann}, and J.~{Staude}, Eds., {\em Astronomical Society of the Pacific Conference Series} {\bf 184}, 309--313  (1999).

\bibitem{2001SoPh..198..197L}
Z.~{Liu} and J.~M. {Beckers}, ``{Comparative Solar Seeing and Scintillation Studies at the Fuxian Lake Solar Station},'' {\em Solar Phys.} {\bf 198}, 197--209  (2001).

\bibitem{2004PASP..116.1143H}
P.~{Hickson} and K.~{Lanzetta}, ``{Measuring Atmospheric Turbulence with a Lunar Scintillometer Array},'' {\em PASP} {\bf 116}, 1143--1152  (2004).

\bibitem{Waldmann:07}
T.~A. Waldmann, T.~Berkefeld, and O.~von~der L\"{u}he, ``Measuring turbulence height profiles using extended sources and a wide-field hartmann-shack wavefront-sensor,'' in {\em Adaptive Optics: Analysis and Methods/Computational Optical Sensing and Imaging/Information Photonics/Signal Recovery and Synthesis Topical Meetings on CD-ROM},  PMA3, Optica Publishing Group  (2007).

\bibitem{2010A&A...513A..25S}
G.~B. Scharmer and T.~I.~M. van Werkhoven, ``S-dimm+ height characterization of day-time seeing using solar granulation,'' {\em A\&A} {\bf 513}, A25  (2010).

\bibitem{2002MNRAS.337..103W}
R.~W. {Wilson}, ``{SLODAR: measuring optical turbulence altitude with a Shack-Hartmann wavefront sensor},'' {\em MNRAS} {\bf 337}, 103--108  (2002).

\bibitem{2012A&A...542A...2K}
A.~{Kellerer}, N.~{Gorceix}, J.~{Marino}, {\em et~al.}, ``{Profiles of the daytime atmospheric turbulence above Big Bear solar observatory},'' {\em A\&A} {\bf 542}, A2  (2012).

\bibitem{2018MNRAS.478.1459W}
Z.~{Wang}, L.~{Zhang}, L.~{Kong}, {\em et~al.}, ``{A modified S-DIMM+: applying additional height grids for characterizing daytime seeing profiles},'' {\em MNRAS} {\bf 478}, 1459--1467  (2018).

\bibitem{2015PASP..127..870R}
D.~{Ren}, G.~{Zhao}, X.~{Zhang}, {\em et~al.}, ``{Multiple-Aperture-Based Solar Seeing Profiler},'' {\em PASP} {\bf 127}, 870  (2015).

\bibitem{2024MNRAS.528.3981R}
X.~{Ran}, L.~{Zhang}, and C.~{Rao}, ``{AC-SLODAR: measuring daytime normalized optical turbulence intensity distribution based on slope autocorrelation},'' {\em MNRAS} {\bf 528}, 3981--3991  (2024).

\bibitem{1975RaSc...10...71F}
D.~L. {Fried}, ``{Differential angle of arrival - Theory, evaluation, and measurement feasibility},'' {\em Radio Science} {\bf 10}, 71--76  (1975).

\bibitem{2023SPIE12638E..12S}
S.~K. {Subramanian} and S.~{Rengaswamy}, ``{Forward modelling of turbulence strength profile estimation using S-DIMM+},'' in {\em Society of Photo-Optical Instrumentation Engineers (SPIE) Conference Series},  A.~{Mahadevan-Jansen}, A.~{Pradhan}, and S.~N. {Unni}, Eds., {\em Society of Photo-Optical Instrumentation Engineers (SPIE) Conference Series} {\bf 12638}, 1263812  (2023).

\bibitem{2004SPIE.5171..219S}
R.~{Sridharan} and A.~R. {Bayanna}, ``{Low-order adaptive optics for the meter aperture solar telescope of Udaipur Solar Observatory},'' in {\em Telescopes and Instrumentation for Solar Astrophysics},  S.~{Fineschi} and M.~A. {Gummin}, Eds., {\em Society of Photo-Optical Instrumentation Engineers (SPIE) Conference Series} {\bf 5171}, 219--230  (2004).

\bibitem{1975OptCo..14..200W}
C.~P. {Wang}, ``{Isoplanicity for imaging through turbulent media},'' {\em Optics Communications} {\bf 14}, 200--204  (1975).

\bibitem{2023aoel.confE..22K}
S.~{Kalyani Subramanian} and S.~{Rengaswamy}, ``{Measurement of isoplanatic angle and turbulence strength profile from H-alpha images of the Sun},'' in {\em Adaptive Optics for Extremely Large Telescopes (AO4ELT7)},  22  (2023).

\bibitem{2004A&A...416.1193A}
A.~{Abahamid}, A.~{Jabiri}, J.~{Vernin}, {\em et~al.}, ``{Optical turbulence modeling in the boundary layer and free atmosphere using instrumented meteorological balloons},'' {\em A\&A} {\bf 416}, 1193--1200  (2004).

\bibitem{2008PASP..120.1318M}
J.~P. {McHugh}, G.~Y. {Jumper}, and M.~{Chun}, ``{Balloon Thermosonde Measurements over Mauna Kea and Comparison with Seeing Measurements},'' {\em PASP} {\bf 120}, 1318  (2008).

\bibitem{2009RaSc...44.2011R}
J.~R. {Roadcap} and P.~{Tracy}, ``{A preliminary comparison of daylit and night C$_{n}$$^{2}$ profiles measured by thermosonde},'' {\em Radio Science} {\bf 44}, RS2011  (2009).

\bibitem{2013arXiv1302.0981N}
A.~{Nayak}, A.~G. {Sreejith}, M.~{Safonova}, {\em et~al.}, ``{High-Altitude Ballooning Program at the Indian Institute of Astrophysics},'' {\em arXiv e-prints} , arXiv:1302.0981  (2013).

\bibitem{2016A&AT...29..397S}
M.~{Safonova}, A.~{Nayaky}, A.~G. {Sreejith}, {\em et~al.}, ``{An overview of high-altitude balloon experiments at the Indian Institute of Astrophysics.},'' {\em Astronomical and Astrophysical Transactions} {\bf 29}, 397--426  (2016).

\bibitem{press_eqn}
D.~{Lide} and H.~{Frederikse}, {\em CRC handbook of chemistry and physics, 1995-1996 : A ready-reference book of chemical and physical data}, CRC Press, Boca Raton, Fl  (1996).

\bibitem{hufnagel}
R.~E. {Hufnagel}, {\em The Infrared Handbook}, Washington, D. C.  (1974).

\bibitem{atst_site_survey}
{{Hill}, F and {Radick}, R and {Collados}, M}, ``{Deriving $C_n^2(h)$ from a Scintillometer Array},'' {\em ATST Project Documentation}   (2003).

\bibitem{kodai_iiap}
S.~{Rengaswamy}, ``{Image Quality Monitoring Experiments at Kodaikanal Tunel Telescope},'' {\em IIA Annual Report 30}   (2016).

\bibitem{2005PASP..117..536A}
M.~{Azouit} and J.~{Vernin}, ``{Optical Turbulence Profiling with Balloons Relevant to Astronomy and Atmospheric Physics},'' {\em PASP} {\bf 117}, 536--543  (2005).

\bibitem{2008JApA...29..345S}
J.~{Singh}, ``{Proposed national large solar telescope},'' {\em Journal of Astrophysics and Astronomy} {\bf 29}, 345--351  (2008).

\end{thebibliography}
\bibliographystyle{spiejour}   


\vspace{2ex}\noindent\textbf{Saraswathi Kalyani Subramanian} is a senior research fellow at the Indian Institute of Astrophysics. She has a B.E. in electronics and communication engineering and an M.Tech in astronomical instrumentation. She is involved in solar AO development at IIA. Her research interest is techniques for achieving high-resolution imaging - AO and interferometry.

\vspace{1ex}

\vspace{2ex}\noindent\textbf{Sridharan Rengaswamy} received Ph.D. in Physics from Bangalore University, India. He worked at the Paranal Observatory (ESO/Chile) as the VLTI Operations astronomer for six years before joining the Indian Institute of Astrophysics as a scientist in 2015. His main fields of research interest are high-resolution imaging, including speckle and interferometric imaging, long baseline interferometry, and adaptive optics. He is also interested in instrumentation, software development for astronomy, and spectroscopy.

\vspace{1ex}

\vspace{2ex}\noindent\textbf{Prasanna Gajanan Deshmukh} is an Engineer-C at the Indian Institute of Astrophysics. After a B.E. in electronics and telecommunication, he specialized in astronomical instrumentation for his M.Tech. and Ph.D. at IIA and the University of Calcutta. His PhD thesis involved the modeling, simulation, and implementation of M1CS for segmented telescopes. His interest lies in telescope control systems, active optics, control system modeling and simulations, astronomical instrumentation, and astronomy outreach.

\vspace{1ex}

\vspace{2ex}\noindent\textbf{Binukumar G. Nair} is a visiting scientist at the Indian Institute of Astrophysics. He was awarded the Marie-Curie fellowship for pursuing a PhD. He received his PhD from The Open University, Milton Keynes, United Kingdom, in 2015. He has worked extensively with UV beamlines in NSRRC, Taiwan, SOLEIL, France, sand ISA, ASTRID, Denmark. His research interests are UV instrumentation, high-altitude ballooning, UV beamlines, and experimental molecular astrophysics

\vspace{1ex}

\vspace{2ex}\noindent\textbf{Mahesh Babu S.} is an electronics engineer (trainee) at the Indian Institute of Astrophysics. He
received his M.Sc. in electronic science from Bangalore University in 2020. He has worked extensively in the electric vehicle manufacturing field as a hardware engineer. His research interests are UV instrumentation, high-altitude ballooning, and electronics hardware development.

\noindent Biographies and photographs of the other authors are not available.

\listoffigures
\listoftables

\end{spacing}
\end{document}